\documentclass[fleqn,usenatbib]{mnras}

\usepackage[T1]{fontenc}

\DeclareRobustCommand{\VAN}[3]{#2}
\let\VANthebibliography\thebibliography
\def\thebibliography{\DeclareRobustCommand{\VAN}[3]{##3}\VANthebibliography}

\usepackage{graphicx}	% Including figure files
\usepackage{amsmath}	% Advanced maths commands
\usepackage{amssymb}	% Extra maths symbols
\usepackage{hyperref}
\usepackage[para,online,flushleft]{threeparttable}

\title[Active asteroid (248370) 2005QN$_{173}$]{Photometric and dynamic characterisation of active asteroid (248370) 2005QN$_{173}$}

\author[B. Novakovi\'c et al.]{Bojan Novakovi\' c$^{1}$%
\thanks{E-mail: \href{mailto:bojan@matf.bg.ac.rs}{bojan@matf.bg.ac.rs (BN)}}%
Debora Pavela$^{1}$
Henry H. Hsieh$^{2,3}$
and Du\v san Mar\v ceta$^{1}$
\\
$^{1}$Department of Astronomy, Faculty of Mathematics, University of Belgrade, Studentski trg 16, 11000 Belgrade, Serbia \\
$^{2}$Planetary Science Institute, 1700 East Fort Lowell Rd., Suite 106, Tucson, AZ 85719, USA\\
$^{3}$Institute of Astronomy and Astrophysics, Academia Sinica, P.O.\ Box 23-141, Taipei 10617, Taiwan}

\date{Accepted XXX. Received YYY; in original form ZZZ}

\pubyear{2022}

\begin{document}
\label{firstpage}
\pagerange{\pageref{firstpage}--\pageref{lastpage}}
\maketitle

\begin{abstract}

We present the physical and dynamical properties of the recently discovered active asteroid (248370) 2005QN$_{173}$ (aka 433P). From our observations, we derived two possible rotation period solutions of $2.7\pm0.1$ and $4.1\pm0.1$~hours. The corresponding light curve amplitudes computed after correcting for the effect of coma are 0.28 and 0.58~mag, respectively. Both period solutions are shorter than the critical rotation limit computed for a strengthless triaxial ellipsoid, suggesting that rotation mass shedding should at least partly be responsible for the observed activity. We confirm that the activity level is fading further, but at a very modest rate of only 0.006~mag/day, still also compatible with sublimation-driven activity. We found that 248370 likely belongs to the Themis asteroid family, making it a fourth main-belt comet associated with this group. Orbital characteristics of 248370 are also consistent with its origin in the young 288P cluster of asteroids. The 288P cluster is associated with its namesake main-belt comet, providing an exciting possibility for a comparative analysis of intriguing main-belt comets 248370 and 288P.

%It should be a single paragraph not more than 250 words (200 words for Letters).

\end{abstract}


%%%%%%%%%%%%%%%%%%%%%%%%%%%%%%%%%%%%%%%%%%%%%%%%%%

%%%%%%%%%%%%%%%%% BODY OF PAPER %%%%%%%%%%%%%%%%%%

\section{Introduction}

Active asteroids are a class of atypical small solar system objects, having at the same time the orbital characteristics of asteroids and the physical characteristics of comets, including coma and tail-like appearance \citep[e.g.][]{2015aste.book..221J}. A subgroup of active asteroids for which it is believed that the sublimation of volatile ices drives observed activity is known as main-belt comets \citep[MBCs;][]{2006Sci...312..561H,2017A&ARv..25....5S}. The MBCs could be a key to tracing the origin and evolution of volatile materials in the asteroid belt and could help our understanding of the protoplanetary disk processes and planetary formation. Though their number has been increasing in recent years, the number of known MBCs is still modest, and there are even fewer of them which are well studied. Any new characterisation of MBCs is therefore important.

The activity of asteroid 248370 was recently discovered by \citet{CBET2021F} in the images collected by the Asteroid-Terrestrial-Impact Last Alert System \citep[ATLAS;][]{2018PASP..130f4505T}. Following this discovery, \citet{2021ApJ...922L...8C} analysed the archival data of 248370 and found that it was active also during its previous perihelion passage in 2016. Based on this recurrent activity \citeauthor{2021ApJ...922L...8C} proposed that activity is sublimation-driven, making asteroid 248370 a main-belt comet. \citet{2021ApJ...922L...9H} further studied the object and performed its physical characterisation. The authors determined various properties of its nucleus, including revised $V$-band absolute magnitude and radius, which are found to be $H_V=16.32 \pm 0.08$~mag and $r_n = 1.6 \pm 0.2$~km, respectively. Based on the archival observations when the object was inactive, the colours of the nucleus and coma are consistent with C-type taxonomic classification. \citeauthor{2021ApJ...922L...9H} also found that activity level is dropping as 248370 moves away from the perihelion, at about 0.01~mag/day, consistent with sublimation-driven activity.

This work presents the additional analysis of the active asteroid (248370) 2005QN$_{173}$ (cometary designation 433P). We performed photometric observations of 248370, aiming primarily to construct its light curve and determine the rotation period, a still missing information to rule out possible mass shedding due to the fast rotation. We also analysed the orbital stability of the object and investigated its possible association with asteroid families.

\section{Observations}

New observations of 248370 were collected on 2021 October 5/6 from the Astronomical station Vidojevica\footnote{\url{http://vidojevica.aob.rs/}} (MPC code C89), using 1.4-m Milankovi\' c telescope. For this purpose, we used Andor iKon-L 2048$\times$2048 pixel CCD camera with a field of view of 13.3$\times$13.3 arcmin and pixel size of 13.5$\times$13.5~$\mu m$. All images were made in standard Johnson-Cousin R-filter, using 2$\times$2 binning. The object was followed over an interval spreading over $\sim$5 hours, and 51 images were taken in total. Additional data on observing circumstances are given in Table~\ref{tab:obs_info}.

\begin{table}
	\centering
	\caption{Observation cicrumstences}
	\label{tab:obs_info}
	\begin{tabular}{lc}
		\hline
		UT Date &  2021 Oct 5/6\\
	    Average seeing (FWHM) & 2.5 arcsec\\
	    Individual exposure times & 200 s \\
	    Total exposure time (stacked image) & 10200 s \\
	    Filter & R \\
	    Heliocentric distance, $r_h$ & 2.491 au \\
	    Geocentric distance, $\Delta$ & 1.518 au \\
	    True anomaly, $\nu$ & 41.9 deg \\
	    Solar phase angle, $\alpha$ & 6.9 deg \\
	    Apperant coma magnitude & 18.51 \\
	    Corresponding absolute magnitude in $R$-band & 15.09\\
	    \hline
	\end{tabular}
\end{table}	    
	    
\subsection{Rotational period}
\label{ss:period}

The main aim of our observational effort was to construct the light curve of active asteroid 248370 and determine its rotational period. On the date of our observations, the object was still active (see Section~\ref{ss:activity}), and therefore nucleus obscuration by coma is expected. Such a situation could make an effort to determine the rotational period to fail, as was the case in a recent example of asteroid Gault. In the case of the asteroid Gault attempts to determine its rotational period while it was still active failed \citep[e.g.][]{2019ApJ...874L..20K,2019ApJ...876L..19J,2020MNRAS.496.2636I}. Only once the asteroid became inactive, its period was successfully obtained \citep{2021ApJ...911L..35P,2021MNRAS.505..245D,2021MNRAS.506.5774C}. It suggests that determining the rotation period of 248370 while it is still active could be challenging. However, probably a key limiting factor complicating the extraction of the rotational period of Gault was the small amplitude ($\Delta m \approx$~0.05~mag) of its light curve \citep{2021ApJ...910L..27L}.
An estimation of the brightness variation in the case of 248370 suggests an amplitude of $\Delta m>$~0.3~mag \citep{2021ApJ...922L...9H}, which could permit a period determination even during an active phase of the object.

\begin{figure}
	\includegraphics[width=\columnwidth]{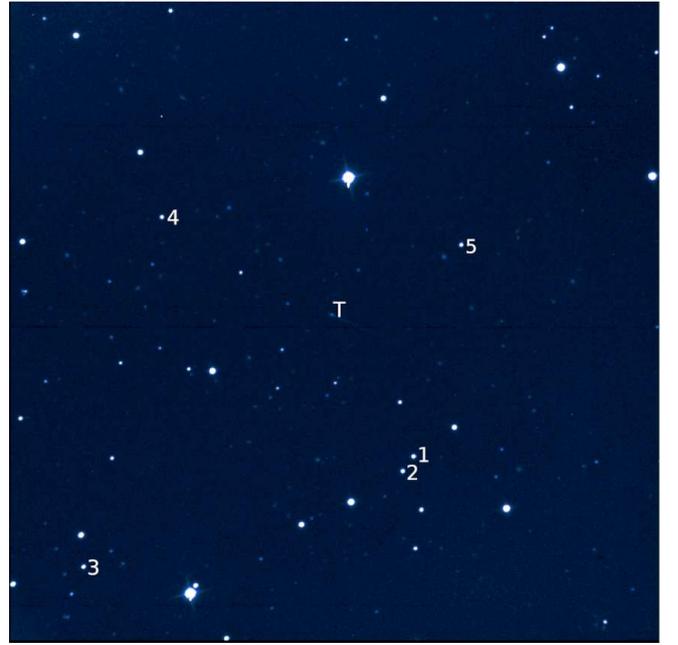}
    \caption{Field of view for one of the images of 248730 active asteroid. The size of the image is 13.3$\times$13.3 arcmin. The target (T) position and five comparison stars (1-5) used to perform photometric measurements are denoted. The image is taken from AS Vidojevica using a 1.4-m Milankovic telescope.}
    \label{fig:FOV}
\end{figure}

For the rotation period determination, image processing, measurement, and period analysis were done using procedures incorporated into the MPO Canopus\footnote{In addition to the period analysis done with MPO Canopus, we also used MATLAB to verify the obtained results additionally. To this purpose, we apply the \emph{fit} function for nonlinear fitting. The option is used to fit the Fourie function to the measurements performed in MPO Canopus and found practically the same rotation period solution as with the FALC routine embedded in MPO Canopus.} \citep[version 10.8.6.3;][]{Warner2021}. The raw images were calibrated with bias, flats, and darks. The $R$-magnitudes were calibrated using the MPOSC3 catalogue\footnote{We have also measured the images using as a reference the ATLAS All-Sky Stellar Reference Catalog \citep{2018ApJ...867..105T}. However, as the obtained results were almost identical and in all cases statistically the same, we kept the values obtained using the MPOSC3 catalogue.}, which converts 2MASS J-K magnitudes to BVRI using formulae developed by \citet{2007MPBu...34..113W}. The Comp Star Selector feature in MPO Canopus was used to limit the comparison stars to near solar colour. The data on comparison starts are listed in Table~\ref{tab:comp_stars}, while their positions in the images are shown in Fig.~\ref{fig:FOV}. 

The period analysis is performed using the Fourier analysis algorithm (FALC) developed by \citet{1989Icar...77..171H}. In particular, the data of apparent magnitude measurements are fitted using the Fourier series of orders 2 and 4. In both cases, this yielded very similar results; therefore, we report only the results obtained using the second-order Fourier series of the form:
\begin{equation}
m - m_0 = \sum_{k=1}^{2} \left( A_k \sin \frac{2 \pi k t}{P}  + B_k \cos \frac{2 \pi k t}{P}  \right) ,
\end{equation}
where $m$ is the observed apparent magnitude, $m_0$ is the light curve midpoint (the average magnitude), $t$ is a time with respect to the starting time, and $A_k$ and $B_k$ are Fourier coefficients.

\begin{figure}
	\includegraphics[width=\columnwidth]{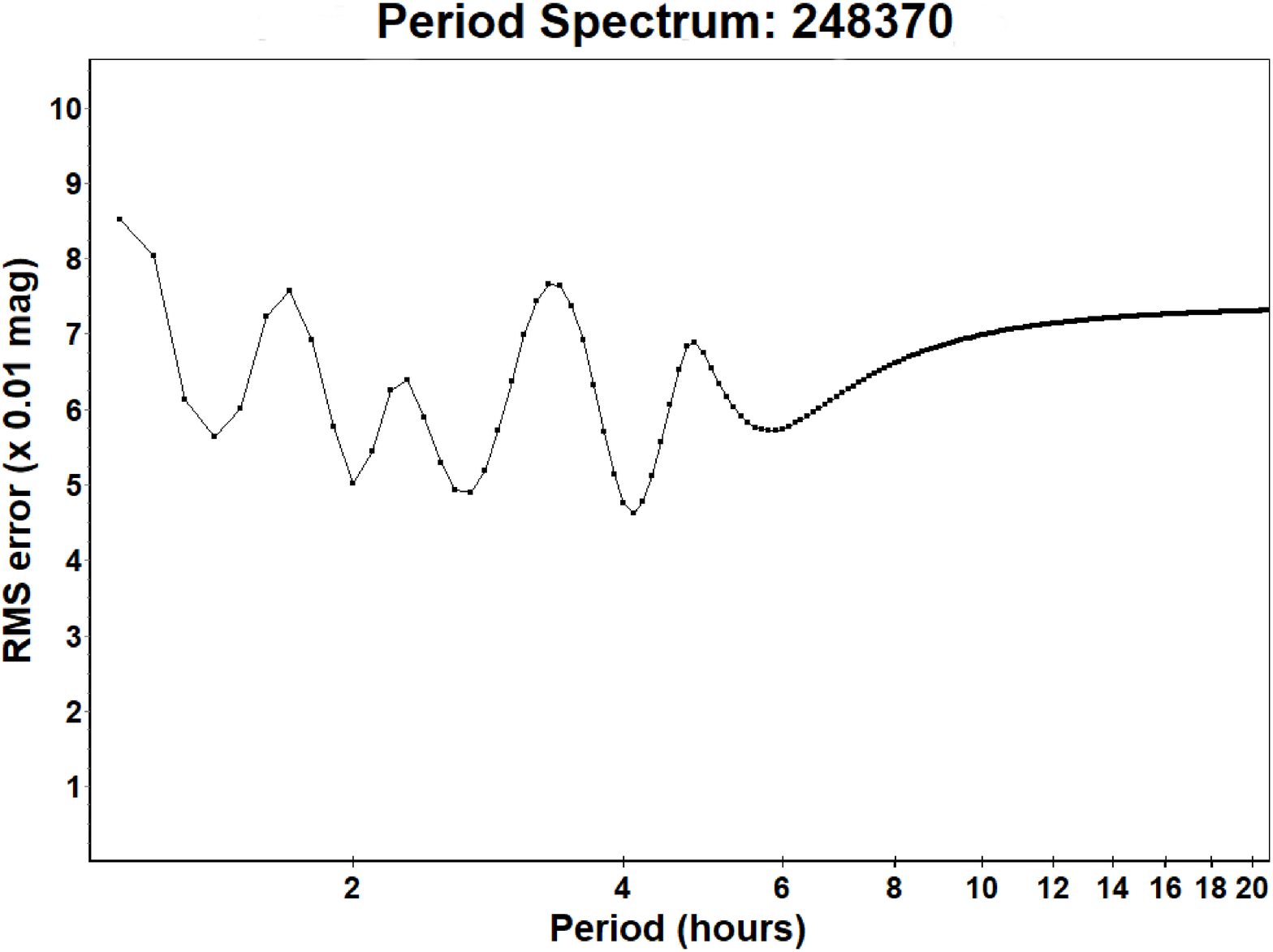}
    \caption{The period spectrum of 248370 from the 2nd order FALC algorithm implemented in MPO Canopus. Three period solutions listed in Table~\ref{tab:periods}, correspond to the three deepest local minima, as indicated in the plot.}
    \label{fig:PeriodSpectrum}
\end{figure}

A periodogram of the period analysis with second-order FALC is shown in Fig.~\ref{fig:PeriodSpectrum}. It suggests there are three potentially realistic period solutions, ranging from about 2 to about 4 hours (Table~\ref{tab:periods}). These solutions all have very similar goodness of fit, with the longest period of P = $4.1\pm0.1$ hours (solution \#3) providing a slightly better fit to the data than the other two. A period of P = $2.0\pm0.1$ hours (solution \#1) corresponds to a single-peaked light curve, i.e. dominated by the first harmonics. However, asteroid light curves are usually dominated by the second harmonic of the rotation period due to the expected elongated shape \citep{2014Icar..235...55H}, which results in a bimodal light curve. Although non-elongated shapes or asteroid obliquity close to 90 degrees (spin axis in the orbit plane) could result in a single-peaked light curve, this situation is less likely, suggesting that our period solution \#1 is less likely than the other two (see the ratio of the amplitudes of the harmonics $H2/H1$ given in Table~\ref{tab:periods}). It also provides a slightly less good fit to the measurements. For these reasons, we discard solution \#1, and consider solutions \#2 and \#3.

\begin{figure}
	\includegraphics[width=\columnwidth]{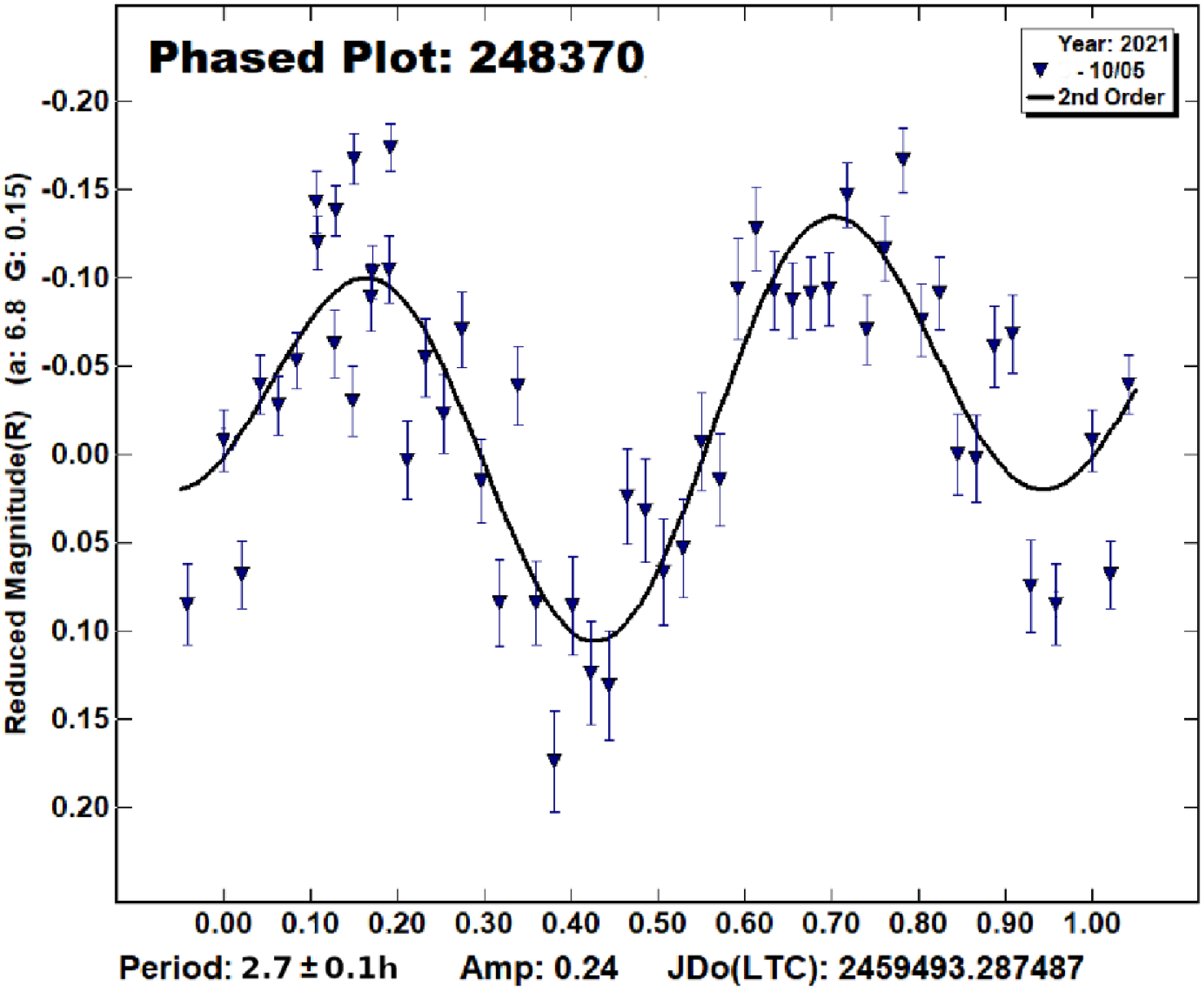}
    \caption{The phased light curves of period solution \#2, plotted with MPO Canopus.}
    \label{fig:PeriodSolution2}
\end{figure}

\begin{figure}
	\includegraphics[width=\columnwidth]{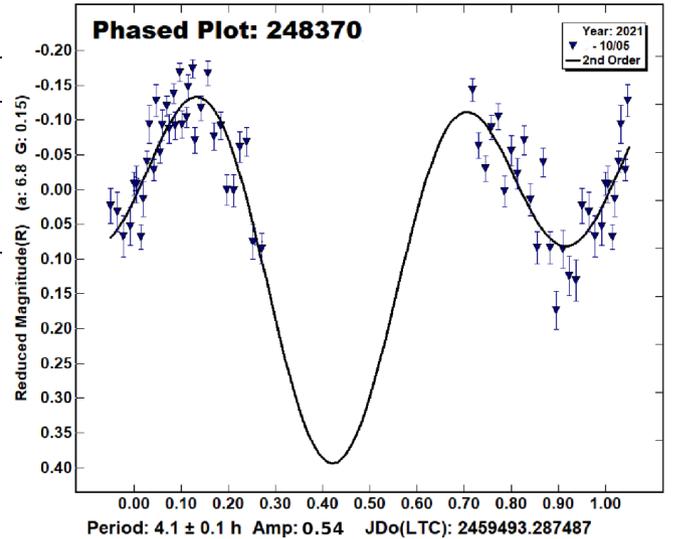}
    \caption{The phased light curves of period solution \#3, plotted with MPO Canopus.}
    \label{fig:PeriodSolution3}
\end{figure}

Solution \#2 has a light curve midpoint at 18.51~mag and an amplitude of 0.24~mag (Fig.~\ref{fig:PeriodSolution2}). A light curve midpoint and an amplitude for solution \#3 are 18.61~mag and 0.54~mag, respectively (Fig.~\ref{fig:PeriodSolution3}). Due to the object's activity, the light curve amplitude is somewhat reduced due to obscuring the nucleus by a coma \citep{2011AJ....142...29H}. Therefore, we first convert our measurements from $R$-band to $r'$-band magnitudes, using
\begin{equation}
r' = R + 0.153 \cdot (r'-i') + 0.117,
\end{equation}
transformation formula from \citet{2006A&A...460..339J}, and the mean $r'-i' =$ 0.1 colour of near-nucleus coma determined by \citet{2021ApJ...922L...9H}. It yields an apparent magnitude of $m_r$=18.64 in the $r'$-band. Combining this result with
an absolute magnitude of the nucleus, determined by \citeauthor{2021ApJ...922L...9H} to be in $r'$-band $H_r=$ 16.12~mag, allows removing the effect of coma on the light curve amplitude. After these corrections, we found that the light curve amplitudes are 0.28 and 0.58~mag for period solutions of $2.7\pm0.1$ (\#2) and $4.1\pm0.1$~hours (\#3), respectively.

Assuming the asteroid is a triaxial ellipsoid of semiaxes $a > b > c$, rotating about the $c$ axis, the $b/a$ axis ratio could be estimated from the relation $\Delta m = 2.5 \log (a/b)$. Using the estimated light curve amplitudes, we found $a/b$ = 1.3 and $a/b$ = 1.7 for the period solutions \#2 and \#3, respectively.\footnote{The axial ratio inferred from the observed light curve amplitude is generally affected by the viewing geometry \citep{2017A&A...598A..91V}. However, as additional observations are needed to account for this effect properly, we neglected it in our calculations.} Both axis ratios are consistent with typical values found for asteroids. Moreover, as 248370 is a member of the Themis family (see Section~\ref{ss:families}), we note that our results are also consistent with the findings by \citet{2008Icar..196..135S}: that is, older families, despite generally tending to contain more spherical-like objects, at larger heliocentric distances tend to have more elongated members.

The critical rotational breakup period, in seconds, of a strengthless ellipsoid, is given by \citet{2018AJ....155..231J} as
\begin{equation}
 P_{crit} = \left ( \frac{a}{b} \right ) \sqrt{ \frac{3\pi}{G \rho} },
 \label{eq:crit_p}
\end{equation}
where $b/a$ is the axial ratio of the ellipsoid, $\rho$ is the density of the asteroid, and $G$=6.674$\times 10^{-11}$~m$^{3}$kg$^{-1}$s$^{-2}$ is the Newtonian gravitational constant.

Assuming a density of $\rho$=1200~kg m$^{-3}$ is appropriate for C-type asteroids \citep{2019Sci...364..268W,lauretta-etal_2019}, and applying Eq.~\ref{eq:crit_p}, we estimated $P_{crit}$ to be $\sim$3.9 and $\sim$5.1 hr, for the period solutions \#2 and \#3 respectively. Interestingly, in both cases, period solutions are shorter than their corresponding critical breakup period, suggesting that rotational instability could play a role in dust ejection.  

A larger density would slightly decrease the critical periods. Also, the obtained values are for strengthless asteroids, though we know that asteroids could have some cohesive strength \citep{2018PEPS....5...25S}. For instance, \citet{2021Icar..36214433Z} found that cohesion is an essential factor in the stability of Didymos. Similarly, \citet{2021A&A...647A..61F} found that a small super-fast rotator 2011PT should have a dust layer at its surface, to explain its small thermal conductivity. However, dust could be kept at the surface of 2011PT only with cohesion. Still, even considering these possibilities, asteroid 248370 rotates fast enough to cause mass sheading. The rotational instability might not be the only driving mechanism for activity, but it likely plays an important role. We will discuss this further in Section~\ref{sec:final}.
 
\begin{table*}
	\centering
	\caption{The basic data on five comparison stars (from MPOSC3 catalogue) used by MPO Canopus for photometric measuremnts and determination of rotational period.}
	\label{tab:comp_stars}
	\begin{tabular}{lcccccccc} % four columns, alignment for each
		\hline
		No &  B & V & R & I & B-V &V-R & $\alpha$& $\delta$ \\
		\hline
		 1  &  17.394 & 16.046 & 15.320 & 14.670 &  1.348  & 0.726 & 23:42:56.85  & -01:53:55.7 \\
		 2  &  17.087 & 16.071 & 15.518 & 15.017 &  1.016  & 0.553 & 23:42:57.75  & -01:54:13.3 \\
		 3  &  16.331 & 15.879 & 15.609 & 15.337 &  0.452  & 0.270 & 23:43:23.63  & -01:56:09.7 \\
		 4  &  17.671 & 16.528 & 15.911 & 15.358 &  1.143  & 0.617 & 23:43:17.40  & -01:49:04.7 \\
		 5  &  17.277 & 16.505 & 16.074 & 15.667 &  0.772  & 0.431 & 23:42:53.04  & -01:49:37.2 \\
		\hline
	\end{tabular}
\end{table*}

\begin{table}
%	\centering
	\caption[]{Three the best-fit rotation period solutions for active asteroid 248370.}
	\begin{threeparttable}
	\begin{tabular}{lccc} % four columns, alignment for each
		\hline
		 &  Solution \#1 & Solution \#2 & Solution \#3 \\
		 \hline
	Period [h]     & 2.0$\pm$0.1 & 2.7$\pm$0.1 & 4.1$\pm$0.1 \\
	Double-peaked  &  No   &  Yes  & Yes   \\
	Mid-point [mag]  & 18.52 & 18.51 & 18.61 \\
	Amplitude [mag]  & 0.23 & 0.24 & 0.54 \\
	H2/H1\tnote{a}   & 0.38 & 1.9 & 1.1 \\
	  \hline  
	\end{tabular}
	\begin{tablenotes}
       \item [a] The ratio of the first to the second harmonic amplitude.
     \end{tablenotes}
  \end{threeparttable}
  	\label{tab:periods}
\end{table}

\subsection{Activity level}
\label{ss:activity}

An essential step in determining the cause of the activity of an active asteroid is monitoring the level of the activity through time. Here we present the measurement of basic activity parameters of 248370, such as length of the tail, coma brightness and $A_d/A_n$ and Af$\rho$ parameters. These values have also been provided by \citet{2021ApJ...922L...9H}, allowing us to compare the values and analyse the activity level over an extended period.

\subsubsection{Length and orientation of the tail}

\begin{figure}
	\includegraphics[width=\columnwidth, angle=0]{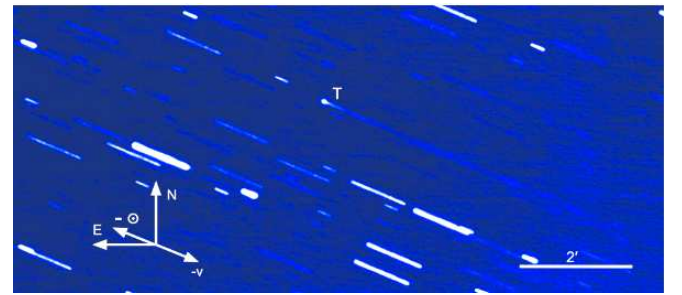}
    \caption{Composite image of 248370 produced by stacking all the 51 images taken at night on 5/6th October 2021, aligned on the object's position. The object's nucleus is located slightly above the centre of the image and is indicated by the letter T. The tail extends down and to the right. The size of the panel is indicated by the scale bar. North (N), East (E), the antisolar ($-\odot$), and the negative heliocentric velocity direction ($-v$) are indicated as well.}
    \label{fig:tail}
\end{figure}

From a stacked image composed of all images taken, we estimated
the length and orientation of the tail of the asteroid 248370. 
The tail projection on the sky extends at least 4 arcmin from the nucleus at a position angle of $247$ degrees (East of North), consistent with the
projected direction of the negative heliocentric velocity vector on the
sky (see Fig.~\ref{fig:tail}). As observations have been done at a geocentric distance of $\Delta$=1.518~au, 4~arcmin correspond to a physical extension of about 264,000~km. This is significantly
shorter than about 720,000~km measured by \citet{2021ApJ...922L...9H}. It could be due to different image depths of the respective stacked images. In this respect, we recall that the tail characteristics are determined in this work from 10200~s of total exposure time from the 1.4-m Milankovic telescope, and \citeauthor{2021ApJ...922L...9H} reports 900~s of total exposure time from 5.1-m Palomar telescope. Doing a pure scaling from aperture size, 900~s on the Palomar corresponds to 11934~s at the 1.4-m telescope. Given that also seeing was similar on the two nights, image depth does not appear to be solely responsible for the change in observed tail length. Therefore, the result could point to a reduced level of activity. Still, we note that observing geometry could also contribute to this change to some degree, as our observations have been made at a lower phase angle of $\alpha$=6.9$^{\circ}$, compared to the observations used by \citeauthor{2021ApJ...922L...9H} which were made on 12 July 2021, at a phase angle of $\alpha$=23.8$^{\circ}$.

We have also checked the expected structure of the tail using the Finson and Probstein theory \citep{1968ApJ...154..327F} implemented in the comet-toolbox\footnote{\url{https://www.comet-toolbox.com/FP.html}} \citep{2014acm..conf..565V}. The tail geometry for the date of our observations suggests that particles ejected more than 35 days prior to the observation date should form a tail aligned with the negative heliocentric velocity vector, as we observed in our images. However, particles released less than 35 days before the observations are expected to produce a tail in the antisolar direction. We note that this is not visible in the images, which might be due to the low activity level in this period.

The slow decline of activity appears to be consistent with sublimation-driven activity on an object moving gradually away from perihelion but inconsistent with activity drivers like impacts, where mass loss is expected to occur in a relatively impulsive event and likely fade quickly.

\subsubsection{Coma brigthness}

\begin{figure}
	\includegraphics[width=\columnwidth, angle=-90]{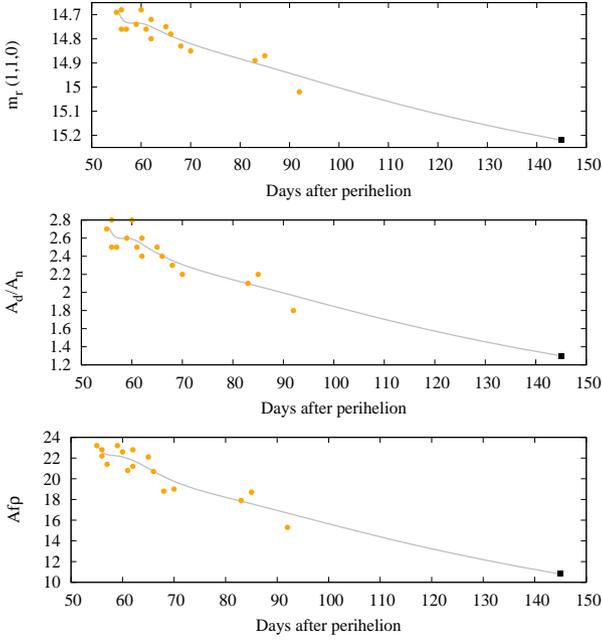}
    \caption{Parameters of 248370's activity level as functions of days after the perihelion passage. Top: $r'$-band magnitude normalized to $r_{h}=\Delta=1$~au, and $\alpha$=0$^{\circ}$; Middle: ratio of dust to nucleus cross-sections; Bottom: Af$\rho$ parameter. Orange circles represent data from \citet{2021ApJ...922L...9H}, while the black square points are our measurement. Grey solid lines represent a bezier fit through the data points, shown as a guide of the activity evolution.}
    \label{fig:activity}
\end{figure}

As discussed in Section~\ref{ss:period}, from photometric measurements, we found a midpoint of our nominal rotational period solution \#2, which also coincides with the average apparent magnitude of our measurements, is $m = 18.51$~mag in $R$-band. It yields an apparent magnitude of $m_r=18.64$ in the sdss $r'$-band, which assuming $G = 0.15$ in the IAU H,G magnitude system \citep{1989aste.conf..524B}, corresponds to an absolute magnitude of $m_r(1,1,0)=15.22$~mag.

According to \citet{2021ApJ...922L...9H}, the absolute magnitude of 248370's bare nucleus is $H_r=16.12 \pm 0.10$~mag.
It implies that in our data taken on 5/6th October 2021, when the object was active, it was about 0.9~mag brighter than expected for a non-active object.

A useful parameter of the activity level is the ratio $A_d/A_n$ of the scattering cross-section of ejected near-nucleus particles 
and the undelaying nucleus. It can be obtained from:
\begin{equation}
 A_d/A_n = \frac{1 - 10^{0.4[m_r(1,1,0)-H_r]}}{10^{0.4[m_r(1,1,0)-H_r]}},
\label{eq:AdAn}
\end{equation}
and we found $A_d/A_n = 1.3 \pm 0.4$. 

In Fig.~\ref{fig:activity}, we plot $m_r(1,1,0)$ and $A_d/A_n$ as functions of time, along with measurements of the same quantities from \citet{2021ApJ...922L...9H}. The results for both parameters suggest clear coma fading, with a rate of
0.006 mag/day, a somewhat slower than 0.01 mag/day estimated by \citeauthor{2021ApJ...922L...9H}.

\subsubsection{Af$\rho$}

The brightness of a cometary coma is proportional to the dust production rate. To deduce the production rate from data acquired under different observational circumstances, the measured brightness must be corrected for all other parameters on which it depends. A standard approach is to use the $A(\alpha = 0^\circ)f\rho$ parameter \citep{1984AJ.....89..579A}. For an ideal steady-state coma, when the dust production rate and velocities of ejected particles are constant, the $A(\alpha = 0^\circ)f\rho$ parameter is independent of aperture radius and can determine the lower limit of the dust production rate \citep{2003PASP..115..981B}. 

It is defined as the ratio between the effective cross-section of comet grains in the field of view and the area of that field itself and can be computed as:
\begin{equation}
 Af\rho = \frac{(2 r_h \Delta)^2}{\rho} \times 10^{0.4[m_{\odot}-m_d(r_h,\Delta,0)]},
\label{eq:afrho}
\end{equation}

\noindent where $\Delta$ is the geocentric distance in cm, $r_h$ is the heliocentric distance in au, $\rho$ is the physical radius in cm of the photometry aperture at the geocentric distance of the comet, and $m_d(r_h, \Delta, 0)$ is the phase-angle-normalized to ($\alpha = 0^\circ$) magnitude of the comet when flux from the nucleus is subtracted. Finally,
$m_{\odot}$ is the Sun’s apparent magnitude for which we used the $r'$-band value of $m_{\odot, r} = -27.05$.

The value of $Af\rho = 10.8 \pm 1.0$ for our measurement shows that about 50 days after \citeauthor{2021ApJ...922L...9H}'s last observation, the activity level has dropped further, consistent with our finding based on the other activity level parameters. Caution is needed to interpret the results for highly asymmetric comae and objects in a non-stationary state. In case of outbursts, or other temporary events that change the activity status of the comet, the $Af\rho$ parameter needs to be interpreted with special care. We also call attention to the fact that the independence of the $Af\rho$ parameter on photometry aperture radius is based on the assumption of spherically symmetric radial outflow \citep{1984AJ.....89..579A}. It may be quite different from the distribution of dust ejected wholly or partially due to an object's fast rotation. In the latter cases, comparing $Af\rho$ values from very different observation epochs and with very different physical aperture sizes may not be particularly meaningful.

\section{Dynamical Characterisation}

\subsection{Long-term dynamical stability}
\label{ss:long-term-dynamics}

Long-term dynamical stability is a key indicator of whether an object possibly spent a very long time at its current location within the main belt or it needs to be transported from elsewhere. Orbital stability over an interval comparable to the age of the solar system or to the age of an associated asteroid family indicates that an asteroid could be formed in situ. Significantly shorter ages imply that it was transported in the past at its current location, not more than the current stability interval ago.

\begin{figure}
	\includegraphics[width=\columnwidth, angle=-90]{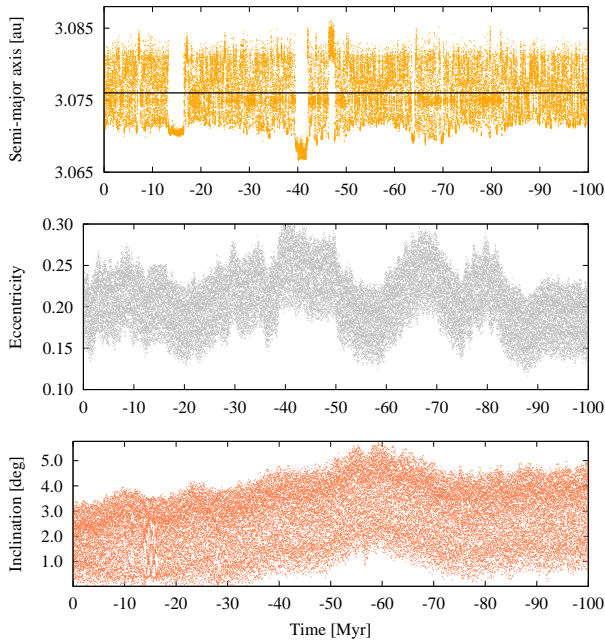}
    \caption{A backward orbit evolution of the nominal orbit of active asteroid 248370 over an interval of 100 Myr. The upper, middle, and bottom panels show the time-evolution of the semi-major axis, eccentricity, and inclination, respectively. The horizontal black line in the upper panel marks the location of the center of the 11:5J resonance.}
    \label{fig:orbit}
\end{figure}

To investigate the long-term dynamical stability of 248370, we first propagated its nominal orbit for 100 Myr. The evolution of the orbital elements is shown in Fig.~\ref{fig:orbit}. The semi-major axis's behaviour suggests that the object is trapped inside an orbital mean-motion resonance, centred around 3.076~au. Further investigation shows that it is 11:5J mean-motion resonance (MMR) with Jupiter. The orbital eccentricity and inclination exhibit a diffusion-like evolution \citep[see][]{nov2010diff}, caused by the interaction with the 11:5J resonance. These results indicate that asteroid 248370 is dynamically an unstable object.

To further assess its stability, we retrieve the proper orbital elements and frequencies \citep{knemil2000}, and Lyapunov time of asteroid 248370 from the Asteroid Families Portal \citep[][see also Appendix A section in \citet{2021A&A...649A.115V}]{nov2022CeMDA}. These data are shown in Table~\ref{tab:pro_ele}. The value of the proper semi-major axis of $a_p = 3.0764$~au confirms that the asteroid interacts with the 11:5J resonance. This is also reflected in its Lyapunov time, which is only $T_L = 4.7$~kyr, suggesting that the past orbit evolution of the object is not deterministic beyond some 50~kyr ago, which is a characteristic of highly unstable orbits.

\begin{table}
	\centering
	\caption{Proper orbital elements: source \href{http://asteroids.matf.bg.ac.rs/fam/}{Asteroid Families Portal}}
	\label{tab:pro_ele}
	\begin{tabular}{lc} % 
		\hline
		Absolute magnitude, H & 15.93 \\
		Semi-major axis, $a_p$ & $3.07604 \pm 0.00100$ [au] \\
		Eccentricity, $e_p$ & $0.17860 \pm 0.00321$ \\
	    Inclination, $i_p$ & $1.25893 \pm 0.06118$ [deg]\\
	    Mean-motion, $n_p$ & $66.71047 \pm 0.03666$ [deg/yr]\\
	    Perihelion frequency, $g$ & $107.57447 \pm 0.32694$ [arcsec/yr]\\
	    Nodal frequency, $s$ & $-101.946258 \pm 1.05180$ [arcsec/yr]\\
	    Lyapunov time, $T_L$ & $4.7$ kyr \\
	    \hline
	\end{tabular}
\end{table}

Therefore, it is unlikely that the object maintains its current orbit for a long time. In principle, it might be evolved recently from some distant parts of the main belt or even beyond. However, this scenario is not likely, particularly considering its very low orbital inclination. Instead, we believe that 248370 entered the 11:5J MMR relatively recently, 
from a nearby region.

We used the orbital and Yarkovsky clones to investigate the dynamical stability further and possible scenarios of asteroid 248370 past orbit evolution. For this purpose, we made 500 clones, and their orbits are propagated for 200 Myr using a version of OrbFit\footnote{Freely available at \url{https://github.com/Fenu24/OrbFit}} software package extended by \citet{FenNov2022} by implementing non-gravitational effects. The dynamical model includes the gravitational effects of the Sun and seven major planets (from Venus to Neptune) as well as non-gravitational forces. The effect of Mercury is taken into account indirectly by applying a barycentric correction to the initial conditions. The clones are treated as massless particles.

The orbit clones are drawn from the multivariate normal distribution, defined by the orbital covariance matrix \citep[e.g.][]{2017ApJ...837L...3M}. The nominal osculating orbital elements and their corresponding uncertainties are taken from the \href{https://ssd.jpl.nasa.gov/tools/sbdb_lookup.html}{JPL Small-Body Database Lookup}. 

To each orbit clone, we associated parameters relevant for the computation of the Yarkovsky and Yarkovsky–O'Keefe–Radzievskii–Paddack (YORP) effects. Based on 248370's colour indices derived by \citet{2021ApJ...922L...9H}, for density and thermal conductivity, we assumed characteristics appropriate for C-type asteroids. Therefore, we picked up these values from normal distributions with mean and standard deviations of $1190 \pm 100$~kg~m$^{-3}$, $0.015 \pm 0.005$~W~m$^{-1}$~K$^{-1}$, respectively, according to recent findings for asteroids Bennu and Ryugu \citep{2019Sci...364..268W,2020SciA....6.3699R,2020Icar..34813835S}. Diameters are drawn from a normal distribution with a mean and standard deviation of $3200 \pm 400$~m, as found for the size of asteroid 248370 by \citet{2021ApJ...922L...9H}. Rotation periods are also chosen from a normal distribution, but allowing two possible solutions that we found in Section~\ref{ss:period}. Finally, obliquities $\gamma$ are chosen randomly between $0$ and $180$ degrees. Additionally, heat capacity is fixed to $C = 600$~J~kg$^{-1}$~K$^{-1}$, based on the measurements for the CI chondrite meteorites \citep{2021JGRE..12607003P}. The other model parameters defining the strength of the YORP effect are set to default values as given in \citet{FenNov2022}.

\begin{figure}
	\includegraphics[width=\columnwidth, angle=0]{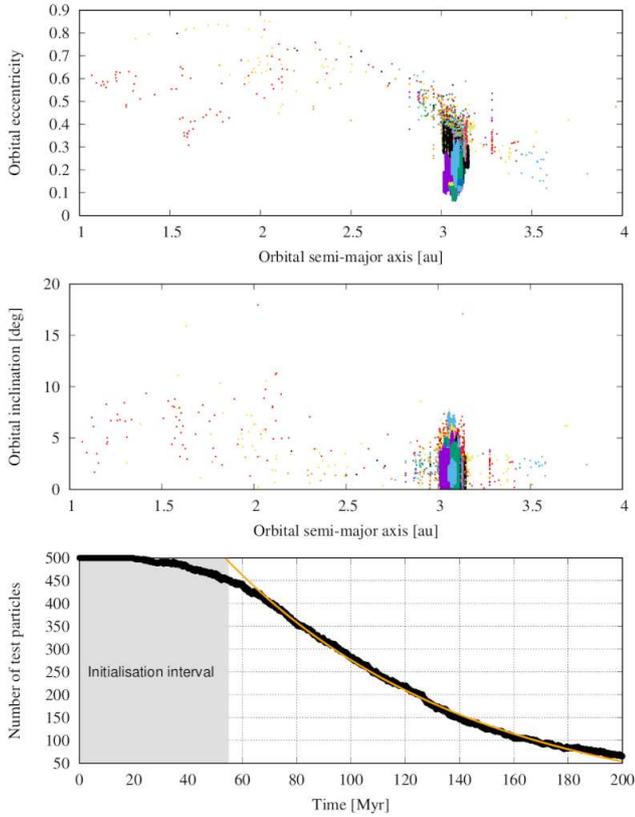}
    \caption{The evolution of 500 test particles inside the 11:5J mean-motion resonance with Jupiter. The test particles represent orbital and yarko clones of asteroid 248370. \em{Upper panel:} The orbital evolution of the test particles over a 200~Myr interval in the semi-major axis vs eccentricity plane. Different colours represent different test particles. \em{Middle panel:} The same as in the upper panel, but in the semi-major axis vs inclination plane \em{Bottom panel:} The black points show the number of surviving test particles inside the resonance as a function of time. After an initialisation interval, needed to fill the resonant region with particles, the resonant population shows an approximately exponential decay in time.
    The orange curve is fit to the data with the function of the form $f(x) = m \cdot exp(-x/t)$, from which we estimated the dynamical half-life of the population to be $\tau \approx 45$~Myr.}
    \label{fig:clones}
\end{figure}

The results of the orbital evolution of the clones are shown in Fig.~\ref{fig:clones}. Under the combined influence of chaotic dynamics and non-gravitational effects, the clones are scattered through the main belt, and some of them even beyond. 
An interesting scenario to note here is that many particles jump from the 11:5J resonance to other nearby mean-motion resonances. This happens because chaotic diffusion increases orbital eccentricity to the point when these resonances start to overlap, allowing an easy switch from resonance to resonance. It implies that an object can survive inside the 11:5J resonance for a relatively limited time. Specifically, from our simulations, we found that the dynamical half-life of the resonant population is $\tau \approx 45$~Myr.

The 11:5J resonance crosses the Themis family, to which 248370 likely also belongs (see Section~\ref{ss:families} and \citep{nov2022CeMDA}). In this respect, \citet{2020AJ....159..179H} found that many former Themis family members that stayed inside the main belt escaped from the family via the 11:5J resonance. Furthermore, the authors found that some of those escaped objects could evolve onto Jupiter-family comets like orbits.

These facts collectively imply that asteroid 248370 is not native to the 11:5J resonance, supporting our hypothesis that it arrived there from a nearby region at some point in the past. In principle, 248370 could either be transported to the resonance by the Yarkovsky effect-induced semi-major axis drift or directly injected inside the resonance by a catastrophic disruption of a member of the Themis family. In the first case, the more likely scenario is that 248370 reached the resonance by drifting inwards because many more family members are located outer to the resonance than inside. For such a scenario, retrograde spin is required, which, assuming the YORP effect did not have enough time to reorient the spin axis, could be observationally tested. To the second scenario, we will come back in Section~\ref{ss:families}.

\subsection{Association to asteroid families}
\label{ss:families}

Many active asteroids are associated with asteroid families, with their subpopulation of main-belt comets being linked to dark carbonaceous groups \citep{2018AJ....155...96H}. Links to families are not relevant only for the water-ice sublimation but also for the constraints on the YORP-cycle timescale relevant for potentially fast-spinning asteroids that shed the mass. For these reasons, we investigated if 248370 is associated with an asteroid family or not.

The standard approach for classifying asteroids into families is the Hierarchical Clustering Method (HCM) proposed by \citet{1990AJ....100.2030Z}. Using the catalogue of proper elements for main-belt asteroids available at the Asteroid Families Portal, we applied the HCM with active asteroid 248370 as a central body \citep[see][for details on this approach]{2017MNRAS.470..576R}. 248370 belongs to the family at a nominal cut-off distance of 55~m~s$^{-1}$. The result suggests that 248370 is a member of the Themis family, which is about 3.3~Gyr old \citep{nov2022CeMDA}.

In order to further test the robustness of our findings, following an approach used by \citet{2018RNAAS...2..129N}, we computed the proper elements for the 100 orbital clones generated in Section~\ref{ss:long-term-dynamics} and applied the HCM to this set of the elements on a case by case basis. For the same cut-off distance of 55~m~s$^{-1}$, only 16 out of 100 cloned orbits are dynamically associated with the Themis family. Increasing the cut-off distance to 60 and 65~m~s$^{-1}$ increases the number of associated clones to 27 and 46, respectively. Though a fraction of the associated clones could seem low, this is expected for highly unstable orbits located at the outskirt of the family (see Fig.~\ref{fig:themis}). In addition to that, as the procedure to compute synthetic proper elements includes the numerical orbit propagation \citep{2017SerAJ.195....1K,2021A&A...649A.115V}, this step increased the orbital eccentricity for the vast majority of the clones, making them less recognizable as the family members. Therefore, though dynamically, 248370 could only marginally be linked to the family, we concluded that it very likely originates in the Themis family.

\subsubsection*{A possible link to the 288P cluster?}
\label{sss:288P-link}

Asteroid 248370 is in the orbital space close to a young sub-family of the Themis family, namely the 288P (aka P/2006VW$_{139}$) cluster of asteroids. The namesake main-belt comet is associated with this about 7.5~Myr old cluster \citep{2012MNRAS.424.1432N}. Given its comet-like behaviour and proximity in the orbital space, 248370 could be potentially somehow related to the 288P main-belt comet and its associated namesake cluster. In the following steps, we examined this possibility.\footnote{For families younger than about 10~Myr, the backward integration method \citep[BIM;][]{karin2002} is an efficient tool to determine their ages, as well as to identify their members \citep{2012MNRAS.424.1432N}. In combination with the classical HCM approach, the BIM could also be very efficient in helping identify new families and their members \citep{rnaas2019nov}. Unfortunately, the BIM cannot be applied due to the dynamical instability of 248370's orbit.}

\begin{figure}
	\includegraphics[width=0.72\columnwidth, angle=-90]{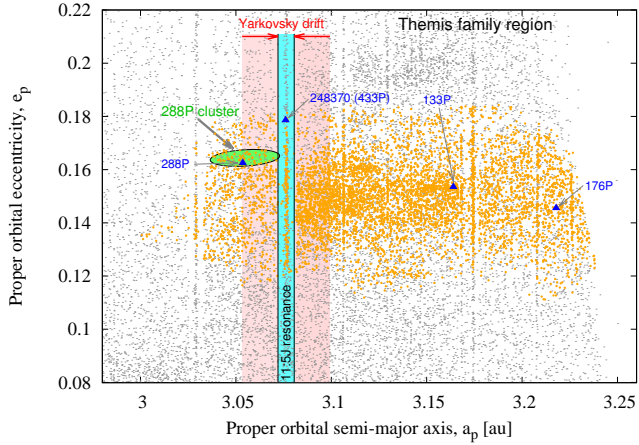}
    \caption{The semi-major axis vs eccentricity plane projection of the Themis family neighbourhood. Grey dots show background asteroids, while orange circles represent Themis family members. Locations of four main-belt comets belonging to the family are shown as blue triangles. Green-shadowed ellipse approximates the initial dispersion of kilometre-sized members of the 288P cluster, and the light-blue area mark width of the 11:5J resonance. The red-shadowed regions on each side of the resonance mark the maximum expected drift in the semi-major axis, due to the Yarkovsky effect, of the asteroid 248370 over 100 Myr.}
    \label{fig:themis}
\end{figure}

\emph{How compatible is the location of 248370 with its possible origin in the 288P cluster?} The barycenter of the cluster in the proper elements space is located at $a_p$=3.0548~au, $e_p$=0.1644 and $i_p$=2.47 degrees \citep{2012MNRAS.424.1432N}. In terms of the orbital eccentricity and inclination, with respect to the centre of the cluster, the object has an offset of 0.0142 and 1.21 degrees, respectively. These differences could well be consequences of the chaotic diffusion inside the 11:5J resonance. An illustrative example is the evolution of 248370's nominal orbit shown in Fig.~\ref{fig:orbit}. Looking at the possible changes in the eccentricity and inclination throughout 10~Myr, we note that they could go up to 0.05 and 1.2 degrees, respectively.

The offset in the proper orbital semi-major axis deserves a more in-depth analysis. A currently observed distance of 248370 from the centre of the 288P cluster is 0.03024~au. Since the object is inside the mean-motion resonance, its distance from the cluster is less relevant. Instead, the distance between the nearer edge of the resonance and the cluster's centre should be considered. With the inner edge of the 11:5J resonance being at about 3.068~au, this distance is about 0.0132~au. If 248370 originates in the 288P cluster, this distance should result from the combined effects of the original impact ejection velocity and the Yarkovsky effect. The maximum Yarkovsky drift for 248370's size over 7.5~Myr (the estimated age of the 288P cluster) should not exceed $\sim 10^{-3}$~au. Therefore, to be compatible with an origin in the impact event forming the 288P cluster, 248370 needs to be ejected 0.011~au from the group's centre.\footnote{Note that in Fig.~\ref{fig:themis} the range of possible Yarkovsky induced drift in the semi-major axis is, for illustration purposes, shown for a longer interval of 100~Myr.} This requires an ejection velocity of about 50~m~s$^{-1}$. 

\emph{How realistic is this ejection velocity in the case of the 288P cluster?}
According to \citet{2010Icar..207...54J}, the median collision ejection velocity $<V_{ej}>$ scales proportionally to the size of the target. As the parent body of the 288P family was 15-20~km in diameter \citep{2012MNRAS.424.1432N}, this velocity\footnote{Estimated for an impact velocity of 3~km~s$^{-1}$ and an impact angle of 45 degrees.} is expected to be about $<V_{ej}>$=50~m~s$^{-1}$ \citep[see Fig. 18 in][]{2010Icar..207...54J}. It is exactly the required velocity for a fragment launched from the 288P cluster's parent body to reach the edge of the 11:5J resonance, implying that 248370 could also be associated with the 288P cluster.

\section{Summary, Discussions and Conclusions}
\label{sec:final}

From the photometric observations, we constructed the light curve of the active asteroid 248370 and found two plausible rotation period solutions. These are $2.7\pm0.1$ and $4.1\pm0.1$~hours. In both cases, the rotation period is shorter than a rotationally-induced mass loss limit, computed for a strengthless ellipsoid. On the other hand, we also found that activity declines consistently with a sublimation-driven activity. It is a hint that both mechanisms, rotational mass shedding and water-ice sublimation, may contribute to the observed activity.

As already discussed by \citet{2021ApJ...922L...9H}, a fast rotation or an elongated shape could reduce the effective gravity felt by dust particles at specific locations on the nucleus surface, allowing them to escape easier. This could lead to a hybrid hypothesis proposed by \citet{2014AJ....147..117J} to explain the activity of the 133P/Elst-Pizarro main-belt comet. In such a scenario, dust launched from the surface by slow gas flow due to sublimating water ice, escapes from the nucleus gravity at possibly sub-escape speeds, assisted by the centripetal acceleration from fast nucleus rotation. Even if the object's rotation period is not strictly shorter than the formally derived critical disruption limit, fast rotation can still lower the effective escape velocity of ejected particles. This could allow those that would otherwise be too slow to escape the nucleus's gravity to be successfully ejected into space instead.

From the dynamical analysis, we found that 248370 is an unstable object, locked inside the 11:5J MMR. Next, we associated it with the Themis family, which is already known as a repository of main-belt comets. More importantly, we also found that dynamical arguments and characteristics are consistent with 248370's origin in a young 288P cluster, a sub-family of the Themis family.

A potential association of 248370 to the 288P cluster is exciting. It suggests possible similarities with 288P comet, which is the only known binary main-belt comet, consisting of two approximately equal size components \citep{2017Natur.549..357A}.
 
The observations provide strong support for sublimation as the activity driver in 288P comet. \citet{2020A&A...643A.152A} favour a scenario of formation and evolution where the 288P binary system formed by rotational splitting following YORP spin-up, and where the activation happened independently of the splitting. Altogether, these findings raised a question about a possible interplay between the binarity, rotational destabilization and sublimation in MBCs \citep[see discussion in][]{nov2022CeMDA}.

Supposing that before the splitting, 288P was about 2-3~km in size, it was only slightly smaller than 248370, whose diameter is between 2.8 and 3.6~km \citep{2021ApJ...922L...9H}. Being active and associated with the 288P cluster and the Themis family, 284370 bears many similarities with 288P main-belt comet. Our findings imply that 248370 rotates close to the rotational disruption limit, and that it is a relatively elongated object. Therefore, its activity could involve multiple contributing driving mechanisms, including sublimation and rotational destabilization. All these facts make 248370 a high-priority object for future studies, as it could provide extremely valuable information to understand better the interplay between the various activity mechanisms in MBCs.

\section*{Acknowledgements}

This work made use of observations obtained with the Milankovic telescope at the Astronomical station Vidojevica (ASV), operated by the Astronomical Observatory of Belgrade. We would like to thank Miodrag Sekuli\' c, telescope operator at ASV, for his help during the observing session. The work of BN and DM has been supported by the Ministry of Education, Science and Technological Development of the Republic of Serbia, contract No. 451-03-68/2022-14/200104. "HHH acknowledges support from NASA via Solar System Observations grant 80NSSC19K0869, NASA Solar System Workings grant 80NSSC17K0723, and Solar System Exploration Research Virtual Institute (SSERVI) Cooperative Agreement grant NNH16ZDA001N"

%%%%%%%%%%%%%%%%%%%%%%%%%%%%%%%%%%%%%%%%%%%%%%%%%%
\section*{Data Availability}

The main data supporting the findings of this study are available within the article. Other data (including raw and observational data) are available on request from the corresponding author.
 
%The inclusion of a Data Availability Statement is a requirement for articles published in MNRAS. Data Availability Statements provide a standardised format for readers to understand the availability of data underlying the research results described in the article. The statement may refer to original data generated in the course of the study or to third-party data analysed in the article. The statement should describe and provide means of access, where possible, by linking to the data or providing the required accession numbers for the relevant databases or DOIs.



\bibliographystyle{mnras}
\bibliography{novakovic_et_al} % if your bibtex file is called example.bib

\begin{thebibliography}{}
\makeatletter
\relax
\def\mn@urlcharsother{\let\do\@makeother \do\$\do\&\do\#\do\^\do\_\do\%\do\~}
\def\mn@doi{\begingroup\mn@urlcharsother \@ifnextchar [ {\mn@doi@}
  {\mn@doi@[]}}
\def\mn@doi@[#1]#2{\def\@tempa{#1}\ifx\@tempa\@empty \href
  {http://dx.doi.org/#2} {doi:#2}\else \href {http://dx.doi.org/#2} {#1}\fi
  \endgroup}
\def\mn@eprint#1#2{\mn@eprint@#1:#2::\@nil}
\def\mn@eprint@arXiv#1{\href {http://arxiv.org/abs/#1} {{\tt arXiv:#1}}}
\def\mn@eprint@dblp#1{\href {http://dblp.uni-trier.de/rec/bibtex/#1.xml}
  {dblp:#1}}
\def\mn@eprint@#1:#2:#3:#4\@nil{\def\@tempa {#1}\def\@tempb {#2}\def\@tempc
  {#3}\ifx \@tempc \@empty \let \@tempc \@tempb \let \@tempb \@tempa \fi \ifx
  \@tempb \@empty \def\@tempb {arXiv}\fi \@ifundefined
  {mn@eprint@\@tempb}{\@tempb:\@tempc}{\expandafter \expandafter \csname
  mn@eprint@\@tempb\endcsname \expandafter{\@tempc}}}

\bibitem[\protect\citeauthoryear{{A'Hearn}, {Schleicher}, {Millis}, {Feldman}
  \& {Thompson}}{{A'Hearn} et~al.}{1984}]{1984AJ.....89..579A}
{A'Hearn} M.~F.,  {Schleicher} D.~G.,  {Millis} R.~L.,  {Feldman} P.~D.,
  {Thompson} D.~T.,  1984, \mn@doi [\aj] {10.1086/113552}, \href
  {https://ui.adsabs.harvard.edu/abs/1984AJ.....89..579A} {89, 579}

\bibitem[\protect\citeauthoryear{{Agarwal}, {Jewitt}, {Mutchler}, {Weaver}  \&
  {Larson}}{{Agarwal} et~al.}{2017}]{2017Natur.549..357A}
{Agarwal} J.,  {Jewitt} D.,  {Mutchler} M.,  {Weaver} H.,   {Larson} S.,  2017,
  \mn@doi [\nat] {10.1038/nature23892}, \href
  {https://ui.adsabs.harvard.edu/abs/2017Natur.549..357A} {549, 357}

\bibitem[\protect\citeauthoryear{{Agarwal}, {Kim}, {Jewitt}, {Mutchler},
  {Weaver}  \& {Larson}}{{Agarwal} et~al.}{2020}]{2020A&A...643A.152A}
{Agarwal} J.,  {Kim} Y.,  {Jewitt} D.,  {Mutchler} M.,  {Weaver} H.,   {Larson}
  S.,  2020, \mn@doi [\aap] {10.1051/0004-6361/202038195}, \href
  {https://ui.adsabs.harvard.edu/abs/2020A&A...643A.152A} {643, A152}

\bibitem[\protect\citeauthoryear{{Bauer}, {Fern{\'a}ndez}  \& {Meech}}{{Bauer}
  et~al.}{2003}]{2003PASP..115..981B}
{Bauer} J.~M.,  {Fern{\'a}ndez} Y.~R.,   {Meech} K.~J.,  2003, \mn@doi [\pasp]
  {10.1086/377012}, \href
  {https://ui.adsabs.harvard.edu/abs/2003PASP..115..981B} {115, 981}

\bibitem[\protect\citeauthoryear{{Bowell}, {Hapke}, {Domingue}, {Lumme},
  {Peltoniemi}  \& {Harris}}{{Bowell} et~al.}{1989}]{1989aste.conf..524B}
{Bowell} E.,  {Hapke} B.,  {Domingue} D.,  {Lumme} K.,  {Peltoniemi} J.,
  {Harris} A.~W.,  1989, in {Binzel} R.~P.,  {Gehrels} T.,   {Matthews} M.~S.,
  eds, Asteroids II. pp 524--556

\bibitem[\protect\citeauthoryear{{Carbognani}, {Buzzoni}  \&
  {Stirpe}}{{Carbognani} et~al.}{2021}]{2021MNRAS.506.5774C}
{Carbognani} A.,  {Buzzoni} A.,   {Stirpe} G.,  2021, \mn@doi [\mnras]
  {10.1093/mnras/stab2111}, \href
  {https://ui.adsabs.harvard.edu/abs/2021MNRAS.506.5774C} {506, 5774}

\bibitem[\protect\citeauthoryear{{Chandler}, {Trujillo}  \& {Hsieh}}{{Chandler}
  et~al.}{2021}]{2021ApJ...922L...8C}
{Chandler} C.~O.,  {Trujillo} C.~A.,   {Hsieh} H.~H.,  2021, \mn@doi [\apjl]
  {10.3847/2041-8213/ac365b}, \href
  {https://ui.adsabs.harvard.edu/abs/2021ApJ...922L...8C} {922, L8}

\bibitem[\protect\citeauthoryear{{Devog{\`e}le} et~al.,}{{Devog{\`e}le}
  et~al.}{2021}]{2021MNRAS.505..245D}
{Devog{\`e}le} M.,  et~al., 2021, \mn@doi [\mnras] {10.1093/mnras/stab1252},
  \href {https://ui.adsabs.harvard.edu/abs/2021MNRAS.505..245D} {505, 245}

\bibitem[\protect\citeauthoryear{{Fenucci} \& {Novakovi\'c}}{{Fenucci} \&
  {Novakovi\'c}}{2022}]{FenNov2022}
{Fenucci} M.,  {Novakovi\'c} B.,  2022, \mn@doi [Serbian Astronomical Journal]
  {10.2298/SAJ2204051F}, \href
  {https://ui.adsabs.harvard.edu/abs/2022SerAJ.204...51F} {204, 51}

\bibitem[\protect\citeauthoryear{{Fenucci}, {Novakovi{\'c}}, {Vokrouhlick{\'y}}
   \& {Weryk}}{{Fenucci} et~al.}{2021}]{2021A&A...647A..61F}
{Fenucci} M.,  {Novakovi{\'c}} B.,  {Vokrouhlick{\'y}} D.,   {Weryk} R.~J.,
  2021, \mn@doi [\aap] {10.1051/0004-6361/202039628}, \href
  {https://ui.adsabs.harvard.edu/abs/2021A&A...647A..61F} {647, A61}

\bibitem[\protect\citeauthoryear{{Finson} \& {Probstein}}{{Finson} \&
  {Probstein}}{1968}]{1968ApJ...154..327F}
{Finson} M.~J.,  {Probstein} R.~F.,  1968, \mn@doi [\apj] {10.1086/149761},
  \href {https://ui.adsabs.harvard.edu/abs/1968ApJ...154..327F} {154, 327}

\bibitem[\protect\citeauthoryear{{Fitzsimmons}, {Erasmus}, {Thirouin}, {Hsieh}
  \& {Green}}{{Fitzsimmons} et~al.}{2021}]{CBET2021F}
{Fitzsimmons} A.,  {Erasmus} N.,  {Thirouin} A.,  {Hsieh} H.~H.,   {Green} D.,
  2021, Central Bureau Electronic Telegrams

\bibitem[\protect\citeauthoryear{{Harris} et~al.,}{{Harris}
  et~al.}{1989}]{1989Icar...77..171H}
{Harris} A.~W.,  et~al., 1989, \mn@doi [\icarus]
  {10.1016/0019-1035(89)90015-8}, \href
  {https://ui.adsabs.harvard.edu/abs/1989Icar...77..171H} {77, 171}

\bibitem[\protect\citeauthoryear{{Harris} et~al.,}{{Harris}
  et~al.}{2014}]{2014Icar..235...55H}
{Harris} A.~W.,  et~al., 2014, \mn@doi [\icarus]
  {10.1016/j.icarus.2014.03.004}, \href
  {https://ui.adsabs.harvard.edu/abs/2014Icar..235...55H} {235, 55}

\bibitem[\protect\citeauthoryear{{Hsieh} \& {Jewitt}}{{Hsieh} \&
  {Jewitt}}{2006}]{2006Sci...312..561H}
{Hsieh} H.~H.,  {Jewitt} D.,  2006, \mn@doi [Science]
  {10.1126/science.1125150}, \href
  {https://ui.adsabs.harvard.edu/abs/2006Sci...312..561H} {312, 561}

\bibitem[\protect\citeauthoryear{{Hsieh}, {Ishiguro}, {Lacerda}  \&
  {Jewitt}}{{Hsieh} et~al.}{2011}]{2011AJ....142...29H}
{Hsieh} H.~H.,  {Ishiguro} M.,  {Lacerda} P.,   {Jewitt} D.,  2011, \mn@doi
  [\aj] {10.1088/0004-6256/142/1/29}, \href
  {https://ui.adsabs.harvard.edu/abs/2011AJ....142...29H} {142, 29}

\bibitem[\protect\citeauthoryear{{Hsieh}, {Novakovi{\'c}}, {Kim}  \&
  {Brasser}}{{Hsieh} et~al.}{2018}]{2018AJ....155...96H}
{Hsieh} H.~H.,  {Novakovi{\'c}} B.,  {Kim} Y.,   {Brasser} R.,  2018, \mn@doi
  [\aj] {10.3847/1538-3881/aaa5a2}, \href
  {https://ui.adsabs.harvard.edu/abs/2018AJ....155...96H} {155, 96}

\bibitem[\protect\citeauthoryear{{Hsieh}, {Novakovi{\'c}}, {Walsh}  \&
  {Sch{\"o}rghofer}}{{Hsieh} et~al.}{2020}]{2020AJ....159..179H}
{Hsieh} H.~H.,  {Novakovi{\'c}} B.,  {Walsh} K.~J.,   {Sch{\"o}rghofer} N.,
  2020, \mn@doi [\aj] {10.3847/1538-3881/ab7899}, \href
  {https://ui.adsabs.harvard.edu/abs/2020AJ....159..179H} {159, 179}

\bibitem[\protect\citeauthoryear{{Hsieh} et~al.,}{{Hsieh}
  et~al.}{2021}]{2021ApJ...922L...9H}
{Hsieh} H.~H.,  et~al., 2021, \mn@doi [\apjl] {10.3847/2041-8213/ac2c62}, \href
  {https://ui.adsabs.harvard.edu/abs/2021ApJ...922L...9H} {922, L9}

\bibitem[\protect\citeauthoryear{{Ivanova}, {Skorov}, {Luk'yanyk}, {Tomko},
  {Hus{\'a}rik}, {Blum}, {Egorov}  \& {Voziakova}}{{Ivanova}
  et~al.}{2020}]{2020MNRAS.496.2636I}
{Ivanova} O.,  {Skorov} Y.,  {Luk'yanyk} I.,  {Tomko} D.,  {Hus{\'a}rik} M.,
  {Blum} J.,  {Egorov} O.,   {Voziakova} O.,  2020, \mn@doi [\mnras]
  {10.1093/mnras/staa1630}, \href
  {https://ui.adsabs.harvard.edu/abs/2020MNRAS.496.2636I} {496, 2636}

\bibitem[\protect\citeauthoryear{{Jewitt}, {Ishiguro}, {Weaver}, {Agarwal},
  {Mutchler}  \& {Larson}}{{Jewitt} et~al.}{2014}]{2014AJ....147..117J}
{Jewitt} D.,  {Ishiguro} M.,  {Weaver} H.,  {Agarwal} J.,  {Mutchler} M.,
  {Larson} S.,  2014, \mn@doi [\aj] {10.1088/0004-6256/147/5/117}, \href
  {https://ui.adsabs.harvard.edu/abs/2014AJ....147..117J} {147, 117}

\bibitem[\protect\citeauthoryear{{Jewitt}, {Hsieh}  \& {Agarwal}}{{Jewitt}
  et~al.}{2015}]{2015aste.book..221J}
{Jewitt} D.,  {Hsieh} H.,   {Agarwal} J.,  2015, in , Asteroids IV.
pp 221--241, \mn@doi{10.2458/azu\_uapress\_9780816532131-ch012}

\bibitem[\protect\citeauthoryear{{Jewitt}, {Weaver}, {Mutchler}, {Li},
  {Agarwal}  \& {Larson}}{{Jewitt} et~al.}{2018}]{2018AJ....155..231J}
{Jewitt} D.,  {Weaver} H.,  {Mutchler} M.,  {Li} J.,  {Agarwal} J.,   {Larson}
  S.,  2018, \mn@doi [\aj] {10.3847/1538-3881/aabdee}, \href
  {https://ui.adsabs.harvard.edu/abs/2018AJ....155..231J} {155, 231}

\bibitem[\protect\citeauthoryear{{Jewitt}, {Kim}, {Luu}, {Rajagopal},
  {Kotulla}, {Ridgway}  \& {Liu}}{{Jewitt} et~al.}{2019}]{2019ApJ...876L..19J}
{Jewitt} D.,  {Kim} Y.,  {Luu} J.,  {Rajagopal} J.,  {Kotulla} R.,  {Ridgway}
  S.,   {Liu} W.,  2019, \mn@doi [\apjl] {10.3847/2041-8213/ab1be8}, \href
  {https://ui.adsabs.harvard.edu/abs/2019ApJ...876L..19J} {876, L19}

\bibitem[\protect\citeauthoryear{{Jordi}, {Grebel}  \& {Ammon}}{{Jordi}
  et~al.}{2006}]{2006A&A...460..339J}
{Jordi} K.,  {Grebel} E.~K.,   {Ammon} K.,  2006, \mn@doi [\aap]
  {10.1051/0004-6361:20066082}, \href
  {https://ui.adsabs.harvard.edu/abs/2006A&A...460..339J} {460, 339}

\bibitem[\protect\citeauthoryear{{Jutzi}, {Michel}, {Benz}  \&
  {Richardson}}{{Jutzi} et~al.}{2010}]{2010Icar..207...54J}
{Jutzi} M.,  {Michel} P.,  {Benz} W.,   {Richardson} D.~C.,  2010, \mn@doi
  [\icarus] {10.1016/j.icarus.2009.11.016}, \href
  {https://ui.adsabs.harvard.edu/abs/2010Icar..207...54J} {207, 54}

\bibitem[\protect\citeauthoryear{{Kleyna} et~al.,}{{Kleyna}
  et~al.}{2019}]{2019ApJ...874L..20K}
{Kleyna} J.~T.,  et~al., 2019, \mn@doi [\apjl] {10.3847/2041-8213/ab0f40},
  \href {https://ui.adsabs.harvard.edu/abs/2019ApJ...874L..20K} {874, L20}

\bibitem[\protect\citeauthoryear{{Kne{\v{z}}evi{\'c}}}{{Kne{\v{z}}evi{\'c}}}{2017}]{2017SerAJ.195....1K}
{Kne{\v{z}}evi{\'c}} Z.,  2017, \mn@doi [Serbian Astronomical Journal]
  {10.2298/SAJ170407005K}, \href
  {https://ui.adsabs.harvard.edu/abs/2017SerAJ.195....1K} {194, 1}

\bibitem[\protect\citeauthoryear{{Kne{\v{z}}evi{\'c}} \&
  {Milani}}{{Kne{\v{z}}evi{\'c}} \& {Milani}}{2000}]{knemil2000}
{Kne{\v{z}}evi{\'c}} Z.,  {Milani} A.,  2000, Celestial Mechanics and Dynamical
  Astronomy, \href {https://ui.adsabs.harvard.edu/abs/2000CeMDA..78...17K} {78,
  17}

\bibitem[\protect\citeauthoryear{{Lauretta} et~al.,}{{Lauretta}
  et~al.}{2019}]{lauretta-etal_2019}
{Lauretta} D.~S.,  et~al., 2019, \mn@doi [\nat] {10.1038/s41586-019-1033-6},
  \href {https://ui.adsabs.harvard.edu/abs/2019Natur.568...55L} {568, 55}

\bibitem[\protect\citeauthoryear{{Luu}, {Jewitt}, {Mutchler}, {Agarwal}, {Kim},
  {Li}  \& {Weaver}}{{Luu} et~al.}{2021}]{2021ApJ...910L..27L}
{Luu} J.~X.,  {Jewitt} D.~C.,  {Mutchler} M.,  {Agarwal} J.,  {Kim} Y.,  {Li}
  J.,   {Weaver} H.,  2021, \mn@doi [\apjl] {10.3847/2041-8213/abedbc}, \href
  {https://ui.adsabs.harvard.edu/abs/2021ApJ...910L..27L} {910, L27}

\bibitem[\protect\citeauthoryear{{Moreno} et~al.,}{{Moreno}
  et~al.}{2017}]{2017ApJ...837L...3M}
{Moreno} F.,  et~al., 2017, \mn@doi [\apjl] {10.3847/2041-8213/aa6036}, \href
  {https://ui.adsabs.harvard.edu/abs/2017ApJ...837L...3M} {837, L3}

\bibitem[\protect\citeauthoryear{{Nesvorn{\'y}}, {Bottke}, {Dones}  \&
  {Levison}}{{Nesvorn{\'y}} et~al.}{2002}]{karin2002}
{Nesvorn{\'y}} D.,  {Bottke} W.~F.,  {Dones} L.,   {Levison} H.~F.,  2002,
  \nat, \href {http://adsabs.harvard.edu/abs/2002Natur.417..720N} {417, 720}

\bibitem[\protect\citeauthoryear{{Novakovi{\'c}}}{{Novakovi{\'c}}}{2018}]{2018RNAAS...2..129N}
{Novakovi{\'c}} B.,  2018, \mn@doi [Research Notes of the American Astronomical
  Society] {10.3847/2515-5172/aad412}, \href
  {https://ui.adsabs.harvard.edu/abs/2018RNAAS...2..129N} {2, 129}

\bibitem[\protect\citeauthoryear{{Novakovi{\'c}} \&
  {Radovi{\'c}}}{{Novakovi{\'c}} \& {Radovi{\'c}}}{2019}]{rnaas2019nov}
{Novakovi{\'c}} B.,  {Radovi{\'c}} V.,  2019, \mn@doi [Research Notes of the
  American Astronomical Society] {10.3847/2515-5172/ab3460}, \href
  {https://ui.adsabs.harvard.edu/abs/2019RNAAS...3..105N} {3, 105}

\bibitem[\protect\citeauthoryear{{Novakovi{\'c}}, {Tsiganis}  \&
  {Kne{\v{z}}evi{\'c}}}{{Novakovi{\'c}} et~al.}{2010}]{nov2010diff}
{Novakovi{\'c}} B.,  {Tsiganis} K.,   {Kne{\v{z}}evi{\'c}} Z.,  2010, \mn@doi
  [\mnras] {10.1111/j.1365-2966.2009.15970.x}, \href
  {https://ui.adsabs.harvard.edu/abs/2010MNRAS.402.1263N} {402, 1263}

\bibitem[\protect\citeauthoryear{{Novakovi{\'c}}, {Hsieh}  \&
  {Cellino}}{{Novakovi{\'c}} et~al.}{2012}]{2012MNRAS.424.1432N}
{Novakovi{\'c}} B.,  {Hsieh} H.~H.,   {Cellino} A.,  2012, \mn@doi [\mnras]
  {10.1111/j.1365-2966.2012.21329.x}, \href
  {https://ui.adsabs.harvard.edu/abs/2012MNRAS.424.1432N} {424, 1432}

\bibitem[\protect\citeauthoryear{{Novakovi{\'c}}, {Vokrouhlick{\'y}}, {Spoto}
  \& {Nesvorn{\'y}}}{{Novakovi{\'c}} et~al.}{2022}]{nov2022CeMDA}
{Novakovi{\'c}} B.,  {Vokrouhlick{\'y}} D.,  {Spoto} F.,   {Nesvorn{\'y}} D.,
  2022, \mn@doi [Celestial Mechanics and Dynamical Astronomy]
  {10.1007/s10569-022-10091-7}, \href
  {https://ui.adsabs.harvard.edu/abs/2022arXiv220506340N} {134, 34}

\bibitem[\protect\citeauthoryear{{Piqueux}, {Vu}, {Bapst}, {Garvie},
  {Choukroun}  \& {Edwards}}{{Piqueux} et~al.}{2021}]{2021JGRE..12607003P}
{Piqueux} S.,  {Vu} T.~H.,  {Bapst} J.,  {Garvie} L. A.~J.,  {Choukroun} M.,
  {Edwards} C.~S.,  2021, \mn@doi [Journal of Geophysical Research (Planets)]
  {10.1029/2021JE007003}, \href
  {https://ui.adsabs.harvard.edu/abs/2021JGRE..12607003P} {126, e07003}

\bibitem[\protect\citeauthoryear{{Purdum} et~al.,}{{Purdum}
  et~al.}{2021}]{2021ApJ...911L..35P}
{Purdum} J.~N.,  et~al., 2021, \mn@doi [\apjl] {10.3847/2041-8213/abf2ca},
  \href {https://ui.adsabs.harvard.edu/abs/2021ApJ...911L..35P} {911, L35}

\bibitem[\protect\citeauthoryear{{Radovi{\'c}}, {Novakovi{\'c}}, {Carruba}  \&
  {Mar{\v{c}}eta}}{{Radovi{\'c}} et~al.}{2017}]{2017MNRAS.470..576R}
{Radovi{\'c}} V.,  {Novakovi{\'c}} B.,  {Carruba} V.,   {Mar{\v{c}}eta} D.,
  2017, \mn@doi [\mnras] {10.1093/mnras/stx1273}, \href
  {https://ui.adsabs.harvard.edu/abs/2017MNRAS.470..576R} {470, 576}

\bibitem[\protect\citeauthoryear{{Rozitis} et~al.,}{{Rozitis}
  et~al.}{2020}]{2020SciA....6.3699R}
{Rozitis} B.,  et~al., 2020, \mn@doi [Science Advances]
  {10.1126/sciadv.abc3699}, \href
  {https://ui.adsabs.harvard.edu/abs/2020SciA....6.3699R} {6, eabc3699}

\bibitem[\protect\citeauthoryear{{Scheeres} \& {S{\'a}nchez}}{{Scheeres} \&
  {S{\'a}nchez}}{2018}]{2018PEPS....5...25S}
{Scheeres} D.~J.,  {S{\'a}nchez} P.,  2018, \mn@doi [Progress in Earth and
  Planetary Science] {10.1186/s40645-018-0182-9}, \href
  {https://ui.adsabs.harvard.edu/abs/2018PEPS....5...25S} {5, 25}

\bibitem[\protect\citeauthoryear{{Shimaki} et~al.,}{{Shimaki}
  et~al.}{2020}]{2020Icar..34813835S}
{Shimaki} Y.,  et~al., 2020, \mn@doi [\icarus] {10.1016/j.icarus.2020.113835},
  \href {https://ui.adsabs.harvard.edu/abs/2020Icar..34813835S} {348, 113835}

\bibitem[\protect\citeauthoryear{{Snodgrass} et~al.,}{{Snodgrass}
  et~al.}{2017}]{2017A&ARv..25....5S}
{Snodgrass} C.,  et~al., 2017, \mn@doi [\aapr] {10.1007/s00159-017-0104-7},
  \href {https://ui.adsabs.harvard.edu/abs/2017A&ARv..25....5S} {25, 5}

\bibitem[\protect\citeauthoryear{{Szab{\'o}} \& {Kiss}}{{Szab{\'o}} \&
  {Kiss}}{2008}]{2008Icar..196..135S}
{Szab{\'o}} G.~M.,  {Kiss} L.~L.,  2008, \mn@doi [\icarus]
  {10.1016/j.icarus.2008.01.019}, \href
  {https://ui.adsabs.harvard.edu/abs/2008Icar..196..135S} {196, 135}

\bibitem[\protect\citeauthoryear{{Tonry} et~al.,}{{Tonry}
  et~al.}{2018a}]{2018PASP..130f4505T}
{Tonry} J.~L.,  et~al., 2018a, \mn@doi [\pasp] {10.1088/1538-3873/aabadf},
  \href {https://ui.adsabs.harvard.edu/abs/2018PASP..130f4505T} {130, 064505}

\bibitem[\protect\citeauthoryear{{Tonry} et~al.,}{{Tonry}
  et~al.}{2018b}]{2018ApJ...867..105T}
{Tonry} J.~L.,  et~al., 2018b, \mn@doi [\apj] {10.3847/1538-4357/aae386}, \href
  {https://ui.adsabs.harvard.edu/abs/2018ApJ...867..105T} {867, 105}

\bibitem[\protect\citeauthoryear{{Vincent}}{{Vincent}}{2014}]{2014acm..conf..565V}
{Vincent} J.,  2014, in {Muinonen} K.,  {Penttil{\"a}} A.,  {Granvik} M.,
  {Virkki} A.,  {Fedorets} G.,  {Wilkman} O.,   {Kohout} T.,  eds, Asteroids,
  Comets, Meteors 2014. p.~565

\bibitem[\protect\citeauthoryear{{Vokrouhlick{\'y}} et~al.,}{{Vokrouhlick{\'y}}
  et~al.}{2017}]{2017A&A...598A..91V}
{Vokrouhlick{\'y}} D.,  et~al., 2017, \mn@doi [\aap]
  {10.1051/0004-6361/201629670}, \href
  {https://ui.adsabs.harvard.edu/abs/2017A&A...598A..91V} {598, A91}

\bibitem[\protect\citeauthoryear{{Vokrouhlick{\'y}}, {Novakovi{\'c}}  \&
  {Nesvorn{\'y}}}{{Vokrouhlick{\'y}} et~al.}{2021}]{2021A&A...649A.115V}
{Vokrouhlick{\'y}} D.,  {Novakovi{\'c}} B.,   {Nesvorn{\'y}} D.,  2021, \mn@doi
  [\aap] {10.1051/0004-6361/202140421}, \href
  {https://ui.adsabs.harvard.edu/abs/2021A&A...649A.115V} {649, A115}

\bibitem[\protect\citeauthoryear{{Warner}}{{Warner}}{2007}]{2007MPBu...34..113W}
{Warner} B.~D.,  2007, Minor Planet Bulletin, \href
  {https://ui.adsabs.harvard.edu/abs/2007MPBu...34..113W} {34, 113}

\bibitem[\protect\citeauthoryear{{Warner}}{{Warner}}{2021}]{Warner2021}
{Warner} B.,  2021, BDW Publishing

\bibitem[\protect\citeauthoryear{{Watanabe} et~al.,}{{Watanabe}
  et~al.}{2019}]{2019Sci...364..268W}
{Watanabe} S.,  et~al., 2019, \mn@doi [Science] {10.1126/science.aav8032},
  \href {https://ui.adsabs.harvard.edu/abs/2019Sci...364..268W} {364, 268}

\bibitem[\protect\citeauthoryear{{Zappala}, {Cellino}, {Farinella}  \& {Kne\v
  zevi\' c}}{{Zappala} et~al.}{1990}]{1990AJ....100.2030Z}
{Zappala} V.,  {Cellino} A.,  {Farinella} P.,   {Kne\v zevi\' c} Z.,  1990,
  \mn@doi [\aj] {10.1086/115658}, \href
  {https://ui.adsabs.harvard.edu/abs/1990AJ....100.2030Z} {100, 2030}

\bibitem[\protect\citeauthoryear{{Zhang}, {Michel}, {Richardson}, {Barnouin},
  {Agrusa}, {Tsiganis}, {Manzoni}  \& {May}}{{Zhang}
  et~al.}{2021}]{2021Icar..36214433Z}
{Zhang} Y.,  {Michel} P.,  {Richardson} D.~C.,  {Barnouin} O.~S.,  {Agrusa}
  H.~F.,  {Tsiganis} K.,  {Manzoni} C.,   {May} B.~H.,  2021, \mn@doi [\icarus]
  {10.1016/j.icarus.2021.114433}, \href
  {https://ui.adsabs.harvard.edu/abs/2021Icar..36214433Z} {362, 114433}

\makeatother
\end{thebibliography}

%\begin{thebibliography}{99}
%\end{thebibliography}

%%%%%%%%%%%%%%%%%%%%%%%%%%%%%%%%%%%%%%%%%%%%%%%%%%

%%%%%%%%%%%%%%%%% APPENDICES %%%%%%%%%%%%%%%%%%%%%

\appendix



%%%%%%%%%%%%%%%%%%%%%%%%%%%%%%%%%%%%%%%%%%%%%%%%%%

% Don't change these lines
\bsp	% typesetting comment
\label{lastpage}
\end{document}